\documentstyle[emulateapj, psfig]{article}

\def\gs{\mathrel{\raise0.35ex\hbox{$\scriptstyle >$}\kern-0.6em
\lower0.40ex\hbox{{$\scriptstyle \sim$}}}}
\def\ls{\mathrel{\raise0.35ex\hbox{$\scriptstyle <$}\kern-0.6em
\lower0.40ex\hbox{{$\scriptstyle \sim$}}}}

\begin{document}

\submitted{Received 1998 -- --; accepted: 1998 -- --}

\lefthead{Blain et al.}
\righthead{Deep Counts of Submillimeter Galaxies}

\title{Deep Counts of Submillimeter Galaxies} 

\author{A.\,W.\ Blain,$\!$\altaffilmark{1,2} J.-P.\ Kneib,$\!$\altaffilmark{2} 
R.\,J.\ Ivison\altaffilmark{3} \& Ian Smail\altaffilmark{4} 
}
\affil{\tiny 1) Cavendish Laboratory, Madingley Road, Cambridge
CB3 0HE, UK}
\affil{\tiny 2) Observatoire Midi-Pyr\'en\'ees, CNRS-UMR5572, 
14 Avenue E.\ Belin, 31400 Toulouse, France}
\affil{\tiny 3) Department of Physics \& Astronomy, 
University College London, Gower Street, London, WC1E 6BT, UK.}
\affil{\tiny 4) Department of Physics, University of Durham, South Road, 
Durham DH1 3LE, UK}
\begin{abstract}
We present the counts of luminous submillimeter (sub-mm) galaxies from
an analysis of our completed survey of the distant Universe seen
through lensing clusters. This survey uses massive cluster lenses with
well-constrained mass models to obtain a magnified view of the background 
sky. This both increases the sensitivity of our sub-mm maps and reduces the 
effects of source confusion.  Accurate lens models are used to correct the 
observed sub-mm source counts for the lens amplification. We show that 
the uncertainties associated with this correction do not dominate the final 
errors. We present sub-mm counts derived from two independent
methods: a direct inversion of the observed sources, which are corrected 
individually for lens amplification, and a Monte Carlo simulation of 
our observations using a parametric model for the background counts, 
which is folded through the lens models and incompleteness estimates to 
determine best-fitting values of the count parameters.  Both methods agree 
well and confirm the robustness of our analysis. 
Detections that are identified with galaxies in the lensing clusters 
in deep optical images 
are removed prior to our analysis, and the results are 
insensitive both to the details of the correction and to the redshift distribution 
of the detections.  We present the 
850\,$\mu$m counts at flux densities between 0.5 and 8\,mJy. The count of 
galaxies brighter than 4\,mJy is $1500 \pm 700$\,deg$^{-2}$, in agreement 
with the value of $2500 \pm 1400$\,deg$^{-2}$ reported by Smail, Ivison, \& 
Blain in 1997. The most accurate 850\,$\mu$m count is determined at 1\,mJy:  
$7900 \pm 3000$\,deg$^{-2}$.  All quoted errors include both Poisson 
and systematic terms.  These are the deepest sub-mm counts published, and they 
are not subject to source confusion because the detected galaxies are separated 
and magnified by the lens.  Down to the 0.5\,mJy limit of our counts, the 
resolved 850\,$\mu$m background radiation intensity is 
$(5 \pm 2) \times 10^{-10}$\,W\,m$^{-2}$\,sr$^{-1}$, comparable to the current 
{\it COBE} estimate of the background. This indicates that the bulk of the 
850\,$\mu$m background radiation originated in distant ultraluminous galaxies. 
\end{abstract}


\section{Introduction} 

The Submillimetre Common-User Bolometer Array (SCUBA; Holland et
al.\ 1999) on the James Clerk Maxwell Telescope (JCMT) has recently
been used to detect a new population of extremely luminous distant
dusty galaxies (Smail, Ivison, \& Blain 1997). By exploiting the
strong magnification bias due to lensing through rich clusters of galaxies, 
we reach fainter detection thresholds in a given observing time, 
while suffering little contamination from cluster galaxies (Blain 1997). 
Consistent numbers of 
sources have also been detected 
in the field
(Barger et al.\ 1998; 
Eales et al.\ 1999; Holland et al.\ 1998; Hughes et al.\ 1998). The results 
from 
these studies have important consequences for the history of star formation 
(Blain et al.\ 1999) and may also provide substantial insight into 
the prevalence of dust-enshrouded active galactic nuclei (AGNs) 
during the early 
evolution of galaxies (Ivison et al.\ 1998).  Smail et al.\ (1997) estimated
counts from six galaxies detected in maps of two clusters, A\,370 and
Cl\,2244$-$02, using simple models of the clusters as singular
isothermal spheres. The complete SCUBA lens survey sample 
includes deeper maps of these two clusters and a further five clusters 
mapped to a sensitivity better than 2\,mJy (Smail et al.\ 1998). 
The analysis of the maps and the source extraction procedures are discussed 
by Smail et al.\ (1997). The full catalog of the SCUBA and follow-up data 
can be found elsewhere (Smail et al.\ 1999).
In this new self-contained analysis of the whole lens survey sample, 
we exploit models of the cluster potentials
derived from deep {\it Hubble Space Telescope} ({\it HST}\,) and
high-resolution ground-based optical imaging and the LENSTOOL
ray-tracing code (Kneib et al.\ 1993) to correct for the effects of lens
amplification and to calculate robust number counts of faint sub-mm galaxies 
below the flux limit of previously published work. The 
calculations are carried out in an Einstein--de Sitter world model, but
the results depend only weakly on the world model assumed.
 
\section{Galaxy counts} 

In a blank field the cumulative sub-mm galaxy counts---the
surface density of galaxies
brighter than a given flux density limit---can be derived by simply
dividing the number of detected sources by the surveyed area, although the 
effects of clustering may have to be considered in small fields. The calculation 
is more complicated in the field of a gravitational lens, which magnifies and 
distorts the background sky---the source plane---by an amount that 
depends on both position and redshift. As a result
of lensing, some regions of the source plane are observed to greater
depths than others, even if the sensitivity across the observed map---the
image plane---is uniform.  With a sufficiently detailed mass model
for the lens, corrections can be applied to model the effects of
lensing accurately, and thus the enhanced sensitivity to 
faint background sources can be exploited.   Moreover, lensing 
provides a further advantage: because sub-mm wave telescopes have
coarse angular resolution as compared with optical telescopes,
source confusion can contribute noise in faint images (Blain, Ivison, \&
Smail 1998; Hughes et al.\ 1998). The flux density of all background sources 
is increased by a lensing cluster, while their mean separation on the sky
is also increased. Both of these features reduce the problem of confusion. 

The lensing effect of the rich clusters exploited in the survey is 
modeled accurately by using multiple-component mass distributions
that describe the extended potential well of the clusters and their more 
massive 
individual member galaxies (e.g.,\ Kneib et al.\ 1996). These
models are derived using the positions of multiply imaged features identified in
high-resolution optical images  and are very well constrained using the
spectroscopic redshifts of these multiple images.  Details of the mass
models employed can be found in Cl\,0024$+$16 (Smail et al.\ 1996),
A\,370 (Kneib et al.\ 1993; B\'ezecourt et al.\ 1999), MS\,0440$+$02 
(Gioia et al.\ 1998), Cl\,0939$+$47 (Seitz et al.\ 1996), A\,1835 (Edge et al.\ 1999), 
A\,2390 (Kneib et al.\ 1999), and Cl\,2244$-$02 (J.-P. Kneib 1998, 
private communication).  
The uncertainties in the magnification of background galaxies derived from 
these models are 10\%--20\% and so are comparable with the 
uncertainty in the absolute calibration of the SCUBA maps. 
 
The lens amplification depends on both the redshift of the lens ($z_l$) and
source ($z_s$), although if $z_s \gg z_l$ this effect is minor.  The
lensing clusters are at redshifts between 0.19 and 0.41, and so the 
effect is small if $z_s \gs 1$.  A complete redshift distribution of 
SCUBA-selected galaxies is gradually being determined (Barger et al.\ 1998,
1999; Hughes et al.\ 1998; Lilly et al.\ 1999; Smail et al.\ 1998). Based
on identifications made in high-quality optical images and extremely
deep radio maps, and follow-up optical spectroscopy (Barger et al.\ 1999), 
we are able to distinguish which SCUBA galaxies are within the clusters, and 
in the foreground and background. The numbers of galaxies assigned to 
each category are listed in Table\,1. It is likely that more than 
80\% of the {\it background} galaxies are at $z_s \ls 5$ (Smail et al.\ 1998), 
and no background candidates have been spectroscopically confirmed at 
$z_s < 0.9$ (Barger et al.\ 1999).
Hence, $z_s \gs 1$ for the background galaxies, and so is indeed expected to be 
much greater than $z_l$. Uncertainties in the redshift distribution of the 
SCUBA galaxies should not affect the derived counts significantly. 
The full sample of spectroscopic identifications of 
foreground/cluster/background galaxies span the redshift range 0.2 to 2.8; 
14 out 
of the 17 detected galaxies have plausible identifications 
(Barger et al.\ 1999). 
The potential systematic errors due to these uncertainties are discussed in 
Section\,3.1. At about the 15\% level, they comprise a small contribution to the 
error budget in the count calculations.

\section{Modeling the Counts}

Two approaches were taken to derive counts from our sub-mm
maps and catalogs:
a direct inversion of the observed source
catalog, and a Monte Carlo method to constrain a parametric model for the
background source counts.  A comparison of the two techniques allows us 
to verify the reliability of the results.

\subsection{Direct Inversion}

The detailed mass models available for the seven clusters observed in
the Smail et al.\ (1998)
sample were used to reconstruct the counts of background
galaxies by correcting for the effects of lensing. Due to the
difficulties of incorporating incompleteness corrections in this
method, we concentrated on the 80\% (greater than 4\,$\sigma$) sample
discussed by Smail et al.\ (1997, 1998). We also conducted 
the analysis using the more liberal 50\% (greater than 
3\,$\sigma$) sample, in order to 
both compare the results 
and assess the systematic error expected from the procedure. SCUBA detections
identified with galaxies within the lensing clusters were excluded from our 
analysis. The number of galaxies identified with cluster sources and  
foreground galaxies are listed in Table\,1. At the faintest flux density, the 50\% 
and 80\% samples contain eight and 14 noncluster galaxies respectively. 

Using the appropriate LENSTOOL models, 
the detected sources identified with background galaxies were mapped from their
observed positions back onto the
source planes at four values of $z_s = 1$, 2, 3, and 4. The flux densities of
the galaxies were also corrected individually for lens amplification, leading 
to true flux densities in the source plane
that are less than the observed values. Again,
we stress that this correction is only weakly dependent on
redshift for our sources at $z_s \gs 1$. None of the background SCUBA sources
can be explained easily as multiple images of the same galaxy; 
however, such multiple images could be detected in deeper integrations. 
The number of galaxies in the catalog that are brighter than a
flux density $S$, $N_{\rm raw}(>S, z)$ is calculated at each redshift
by simply counting the number of sources brighter than $S$ {\em in the
source plane, after correcting for lensing}. A simple Poisson
uncertainty is attached to this value.
 
\vskip 2mm 
\hbox{~}

\centerline{\psfig{file=fig1.ps,width=3.3in,angle=270}}
\noindent{
\scriptsize \addtolength{\baselineskip}{-3pt} 
\vskip 1mm
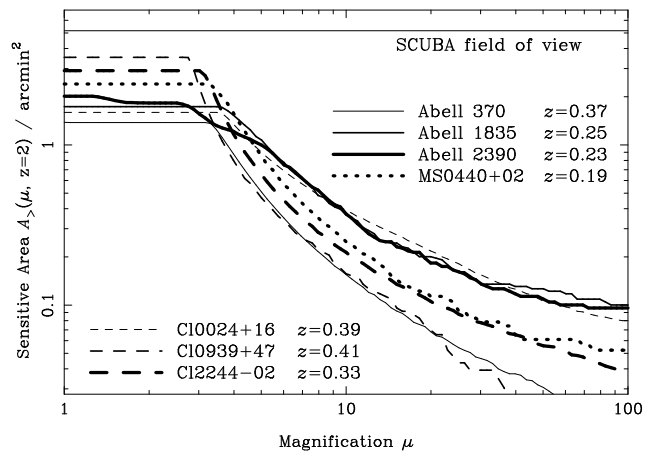
Fig.~1.\ The cumulative area of the source plane $A_>$, at $z_s=2$, that 
experiences a magnification greater than $\mu$ for all seven clusters in the 
Smail et al.\ (1998) survey. The differences between clusters reflect their different 
redshifts and mass distributions. 
Low magnifications are absent because
the SCUBA field of view 
encompasses only 
the concentrated cores
of the clusters. 

\vskip 3mm
\addtolength{\baselineskip}{3pt}
}

The area of background sky within which a galaxy would be detectable in
each cluster was determined from a map of the magnification in the
source plane, derived using the LENSTOOL models. This
quantity is also a weak function of redshift for $z_s \gs 1$. The area
of the source plane behind each cluster that lies within the SCUBA
field of view and is magnified by a factor greater than $\mu$, $A_>$,
is shown in Figure\,1, as an example for $z_s=2$. Due to the
magnification, $A_>$ is smaller than the SCUBA field of view. 

A galaxy with a flux density $S$ in the source plane will appear
in the image plane of a particular cluster above a detection
threshold $S_{\rm min}$ if it is magnified by a factor greater than
$\mu = S_{\rm min}/S$. The area in the source plane within which such a 
galaxy would be detected in that cluster is thus $A_>(S_{\rm min}/S, z)$. The
flux density threshold $S_{\rm min}$ and the form of $A_>$ are different for 
each cluster.  By dividing the number of detected galaxies $N_{\rm raw}$ by
the sum of the areas $A_>(S_{\rm min}/S, z)$ for all seven clusters,
the cumulative count 
$N(>S, z) \simeq N_{\rm raw}(>S, z) / \sum A_>(S_{\rm min}/S, z)$ is found.
The count of identified foreground field galaxies is then 
calculated in the normal way and added to the result. 

The count was calculated independently using this method for 
redshifts $z_s=1$, 2, 3 and, 4. The spread in the results between these four 
cases as a function of flux density threshold is shown in Figure\,2. The results 
obtained from both the 50\% and 80\% samples at each redshift are compared 
with the mean count in each sample. The counts are remarkably consistent, 
reflecting partly that the value of the power-law index $\gamma$ in the count 
$N(> S) \propto S^\gamma$ is $\gamma \ls -1$, 
indicating a small but positive magnification bias. The redshift-dependent 
scatter within the results obtained using both the 50\% and 80\% samples 
is never more than 30\%. Averaged over the 
range of flux densities from 0.5 to 8\,mJy, the mean scatter is 12\%.
Providing that $z_s \gs 1$ for background galaxies, as suggested by the initial 
spectroscopic results (Barger et al.\ 1999),  the systematic uncertainty in the
results due to the redshift distribution of the detected background galaxies is
expected to be smaller than the absolute calibration uncertainties in the 
SCUBA images. These systematic uncertainties are included in the reported
errors.   

To reinforce this point, if the galaxies that are identified in the foreground of 
the clusters are assumed to be mis-identified, and are actually background 
galaxies, then the counts at $S<2$\,mJy are expected to increase by about 
0.5$\sigma$ with the brighter counts remaining the same. If all the background 
galaxies are assumed to be at $z_s=4$, then the count at flux densities less 
than and greater than 2\,mJy is expected to be reduced and increased by about 
0.2$\sigma$, respectively. 

\vskip 2mm
\hbox{~}

\centerline{\psfig{file=fig2.ps,width=3.3in,angle=270}}
\noindent{
\scriptsize \addtolength{\baselineskip}{-3pt} 
\vskip 1mm
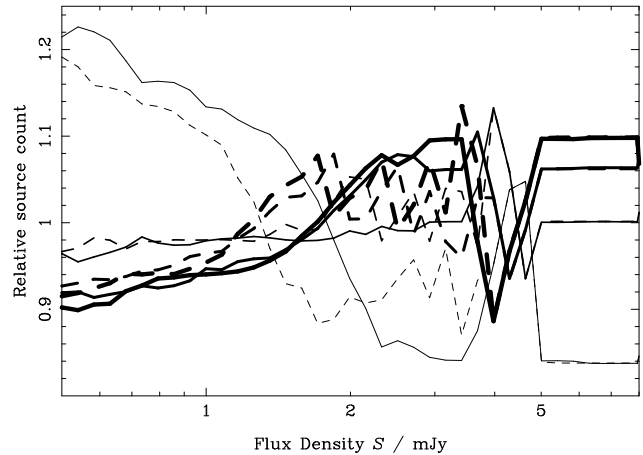
Fig.~2.\ Systematic differences in the inferred counts for the 80\%
Smail et al.\ (1998) sample ({\it solid lines}) 
for different assumed median redshifts 
for the sources.  The line thickness increases in the order $z_s=1$, 2, 3, 
and 4.
We also show the corresponding result for the 50\% sample ({\it dashed lines}).
The count values are normalized to the mean count derived from all four 
redshift values. These is little variation in the counts.  
At worst the counts for $z_s=1$ and $z_s=4$ differ by 30\%. The mean 
scatter of the results about the mean for the four different redshifts 
between flux densities of 0.5 and 8\,mJy is 
12\% in both the 50\% and 80\% samples. Note for comparison that the
Poisson error on the counts is about 35\%. 

\vskip 3mm
\addtolength{\baselineskip}{3pt}
}

\subsection{Parametric Simulations}

To confirm the reliability of the results from the direct inversion we
performed Monte Carlo simulations of our observations using a
parametric model for the background galaxy counts, and thus derived the 
best-fit parameters by comparison with the observations.  This
technique has the advantage that we can explicitly include the
incompleteness of our maps at faint flux densities and so derive the counts 
separately for both the 80\% and 50\% samples. Galaxies were drawn
from a population with a count described by a power-law model:
$N(>S) = K (S/S_0)^\alpha$.  For each realization we selected a value of
$K$ and $\alpha$ and populated the source planes of the seven clusters at 
random for each of the four values of redshift used
above.  The source planes were then mapped onto the image planes using
the LENSTOOL models. The source catalogs derived from the areas of the
image planes for each cluster covered by our SCUBA observations were then
convolved with the appropriate incompleteness function determined by 
Smail et al.\ (1998) to determine the number of galaxies that would be
detected in each cluster in the 50\% and 80\% samples. One thousand 
realizations of this process were executed for each set of model parameters 
[$K$,$\alpha$] to derive a Monte Carlo estimate of the count of background 
galaxies.  As for the direct method, counts were calculated assuming 
$z_s = 1$, 2, 3, and 4 for the background galaxies. The predicted counts 
from each cluster and value of $z_s$ were then added and compared with 
the observed counts for all seven clusters, assuming Poisson 
statistics, and the probability that the observed counts could be produced 
from each [$K$,$\alpha$] pair was determined. The most probable values
of $K$ and $\alpha$ were determined at flux densities $S_0=1$, 2, and 4\,mJy.

\begin{table*}[hbt]
{\scriptsize
\begin{center}
\centerline{Cumulative 850\,$\mu$m counts}
\vspace{0.1cm}
\noindent{
\scriptsize \addtolength{\baselineskip}{-3pt} 
\vskip 0.5mm
%\centerline{
{\sc Note}---
Deduced from both the direct
analysis of the 80\% sample and the Monte Carlo (MC) analysis of both the
80\% and 50\% samples. Note that the errors on the counts are not
independent and that they should not be used to derive differential counts.
The number of galaxies in the 80\% and 50\% samples that are brighter than
each flux density, measured in the source plane, are given
in the fifth and sixth columns, in the form of numbers of
background/foreground/cluster galaxies. \hfill
%}
\vskip 3mm
\addtolength{\baselineskip}{3pt}
}
%The 850-$\mu$m counts deduced from both the direct and 
%Monte-Carlo analyses. The number of galaxies in the 80 and 50\% samples
%at each flux density}
%\centerline{are given 
%in the second and third columns, in the form of numbers of 
%background/foreground/cluster galaxies.} 
%\vspace{0.3cm}
\begin{tabular}{cccccc}
\hline\hline
\noalign{\smallskip}
Flux & Direct & MC $K$ & MC $\alpha$ 
& $N_{80\%}$ & $N_{50\%}$ \cr
(mJy) & (10$^3$ deg$^{-2}$) & (10$^3$ deg$^{-2}$) &  
& & \cr

\hline
\noalign{\smallskip}
0.5 & $22 \pm 9$ & ... & ... & 7/1/2 & 12/2/3 \cr
1.0 & $8.0 \pm 3.0$ & $8.6 \pm 3.0$ & $-1.3 \pm 0.4$ & 7/1/2 & 11/2/3 \cr
2.0 & $2.6 \pm 1.0$ & $3.9 \pm 1.3$ & $-1.4 \pm 1.0$ & 7/1/2 & 11/2/3 \cr
4.0 & $1.5 \pm 0.7$ & $1.4 \pm 0.5$ & $-1.8 \pm 0.7$ & 6/1/2 & 6/2/3 \cr
8.0 & $0.8 \pm 0.6$ & ... & ... & 3/1/1 & 3/1/1 \cr
\noalign{\smallskip}
\noalign{\hrule}
\noalign{\smallskip}
\end{tabular}
\end{center}
}
\vspace*{-0.8cm}
\end{table*}

\vskip 2mm
\hbox{~}

\centerline{\psfig{file=fig3.ps,width=3.3in,angle=270}}
\noindent{
\scriptsize \addtolength{\baselineskip}{-3pt} 
\vskip 1mm
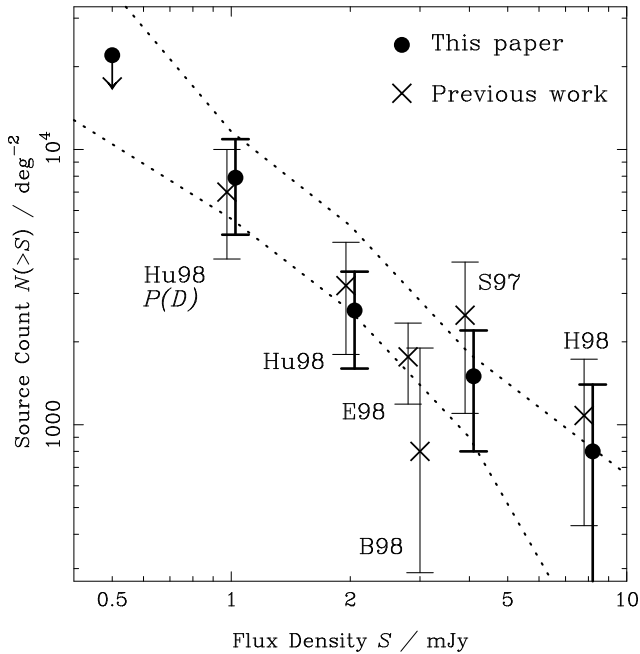
Fig.~3.\ Cumulative sub-mm counts corrected for lens magnification as 
listed in Table\,1. The results obtained by direct inversion are 
shown by solid circles, while the $\pm 1\sigma$ bounds to the counts  
estimated from Monte Carlo simulations are shown by dotted lines.  
Earlier counts from Barger et al.\ (1998; B98), Eales et al.\ (1999; E98),
Holland et al.\ (1998; H98), Hughes et al.\ (1998; Hu98) and Smail et al.\
(1997; S97) are shown by crosses. $P(D)$ signifies the result of Hughes 
et al.'s confusion analysis. 

\vskip 3mm
\addtolength{\baselineskip}{3pt}
}

\section{Results} 

The counts determined using both the direct and Monte Carlo methods are 
listed in Table\,1 and shown in Figure\,3.   
For both methods, the fractional errors obtained when the counts were 
calculated assuming redshifts $z_s=1$, 2, 3, and 4 were very similar. 
The mean of these errors was used to define the final random error on the 
count. The quoted errors also include a systematic term, equal to the fractional 
spread of the results obtained for the 50\% and 80\% samples and for all four 
values of $z_s$. The errors reported in Table\,1 are thus the most conservative 
that can be formulated from the data. In order to be 
maximally conservative, the faintest count at 0.5\,mJy is plotted in Figure\,3, at
the value listed in Table\,1, but as an upper limit rather than a 
detection. It is 
possible for the counts to converge at a flux density just less than 1\,mJy, but 
still remain consistent with the observed catalogs. 

The direct and parametric estimates of the faint sub-mm
counts in Table\,1 and Figure\,3 are fully consistent, supporting 
the reliability and robustness of these results. 
The new counts 
are deeper and more accurate than 
any existing data, including the 
fluctuation analysis of the extremely deep 
SCUBA image of the Hubble Deep Field (Hughes et al.\ 1998). 
This shows the advantages of making sub-mm 
observations of cluster lenses for which accurate mass models are available.
At brighter 
flux density levels, the results are coincident with and have comparable 
errors to the initial 
results of the SCUBA survey in Canada--France Redshift Survey 
fields (Eales et al.\ 1999). 

By integrating over the counts brighter than 0.5\,mJy, a 850\,$\mu$m
background radiation intensity of 
$(5 \pm 2) \times 10^{-10}$\,W\,m$^{-2}$\,sr$^{-1}$ in resolved sources is 
obtained. This is comparable to the current background radiation intensity 
obtained from {\it COBE} data by Fixsen et al.\ (1998). 

\section{Conclusions} 

By exploiting the magnification due to clusters of galaxies, we have determined 
the surface density of submillimeter-luminous 
galaxies down to a very faint flux 
density limit of 0.5\,mJy at 850\,$\mu$m, several times fainter than the confusion 
limit of the JCMT in blank fields. The results are consistent with those from an
initial analysis of a subset of our catalog (Smail et al.\ 1997) and 
with the shallower counts from blank-field surveys. The errors on the
counts are now reduced by a factor of 2 as compared with our earlier 
work. The count of 850\,$\mu$m galaxies brighter than 1\,mJy is 
$7900 \pm 3000$\,deg$^{-2}$. We have shown that the systematic 
errors due to both the uncertainties in the redshift distribution of the 
detected galaxies and the mass models of the clusters studied are less than 
25\%, and thus comparable to the absolute flux calibration of
the SCUBA maps.

By summing the new counts down to their flux density limit, we 
have shown that our SCUBA maps resolve the major part of the 850\,$\mu$m 
background radiation intensity into individual ultraluminous galaxies with 
luminosities greater than $10^{12}\,L_\odot$ (equivalent to star-formation 
rates greater than $100\,M_\odot$\,yr$^{-1}$ if no AGN contribution is 
present). Hence, 
most of the dust-enshrouded star-formation/AGN activity in the distant 
Universe took place in rare but extreme events.
  
Deeper SCUBA observations in the fields of rich lensing clusters will allow
the counts ultimately to be pushed to even fainter flux densities than
those achieved here, especially if an upgraded SCUBA becomes available.  
However, a significant improvement of our knowledge of the faint 
sub-mm-wave counts awaits the commissioning of a large 
millimeter-wave interferometer array.

\section*{Acknowledgements}

We thank 
Malcolm Longair and an anonymous referee 
for helpful comments on the manuscript, 
the SCUBA commissioning team and Ian Robson for their help
throughout the SCUBA lens survey,  
and the PPARC (R. J. I.), Royal Society (I. S.) and MENRT (A. W. B.) 
for support.  
This research has been conducted under a European 
TMR network with generous support 
from the European Commission.

\end{document}